\documentclass[journal]{IEEEtran}
\usepackage{epsfig,graphics,epsf,color,amsmath,balance,cite,amsfonts,multirow,stfloats,algorithm,algorithmic,mathrsfs,amssymb}
\usepackage{mathtools}

\newtheorem{remark}{Remark}
\newtheorem{proposition}{Proposition}

\title{Beamforming Design with Fast Convergence for IRS-Aided Full-Duplex Communication}

\author{Hong~Shen,~\IEEEmembership{Member,~IEEE,}~Tian~Ding,~Wei~Xu,~\IEEEmembership{Senior~Member,~IEEE,}~and~Chunming~Zhao,~\IEEEmembership{Member,~IEEE}\thanks{ This work was supported by the National Natural Science Foundation of China under Grants 61871108 and 61871109, and the Natural Science Foundation of Jiangsu Province for Distinguished Young Scholars under Grant BK20190012. The authors are with the National Mobile Communications Research Laboratory, Southeast University, Nanjing 210096, China (e-mail: \{shhseu, dingtian, wxu, cmzhao\}@seu.edu.cn). W. Xu is also with the Purple Mountain Laboratories, Nanjing 211111, China. \emph{(Corresponding authors: Hong Shen; Wei Xu.)}}}

\begin{document}

\maketitle

\begin{abstract}
We study the beamforming optimization for an intelligent reflecting surface (IRS)-aided full-duplex (FD) communication system in this letter. Specifically, we maximize the sum rate of bi-directional transmissions by jointly optimizing the transmit beamforming and the beamforming of the IRS reflection. A fast converging alternating algorithm is developed to tackle this problem. In each iteration of the proposed algorithm, the solutions to the transmit beamforming  and the IRS reflect beamforming are obtained in a semi-closed form and a closed form, respectively. Compared to an existing method based on the Arimoto-Blahut algorithm, the proposed method achieves almost the same performance while enjoying much faster convergence and lower computational complexity.
\end{abstract}


\section{Introduction}
Intelligent reflecting surface (IRS) assisted wireless communications have recently attracted a plethora of research interests  \cite{Liaskos2018CM,Wu2020CM}. Typically, IRS is composed of a number of low-cost reflecting elements whose amplitudes and phase shifts can be flexibly tuned to fulfill various requirements, e.g., enhancing the signal strength, mitigating the interference, or improving the secrecy.

In traditional communication systems, the transmitter and the receiver usually work under the half-duplex (HD) mode. Therefore, the uplink and downlink transmissions are separated in either a time-division duplex (TDD) or a frequency-division duplex (FDD) manner. In order to further improve the system spectral efficiency, the innovative full-duplex (FD) techniques have been advocated such that the uplink and downlink share the same time-frequency resources {\cite{Kim2015CST,Zhang2016Proc,Khalili2020Arxiv,Khalili2019Conf,Aslani2019TVT}}.


Regarding various IRS-aided HD systems, there have been many works focusing on the joint optimization of transmit beamforming and IRS reflect beamforming, i.e., phase shift matrix. For instance,  beamforming designs for single-user multiple-input
single-output (MISO) systems have been concerned in \cite{Wu2018GLOBECCOM}. The generalization to the multiuser MISO case was studied in {\cite{Wu2019TWC,Huang2019TWC,Zhao2020Arxiv}}. Moreover, the authors of \cite{Hong2019CL} and \cite{Cui2019WCL} investigated the joint beamforming optimization for IRS-aided systems from the perspective of enhancing physical-layer secrecy. Alternatively, the beamforming design for an IRS-assisted simultaneous wireless information and power transfer (SWIPT) system was studied in \cite{Wu2019WCL}.

To the best of our knowledge, the IRS-aided FD system has rarely been considered, except for a few works \cite{Zhang2020CL,Xu2020Arxiv}. It turns out that the corresponding transmission optimization problem is quite hard even for the point-to-point system \cite{Zhang2020CL}. More specifically, concerning the problem of sum rate maximization for the IRS-aided FD multiple-input
multiple-output (MIMO) system, the authors of \cite{Zhang2020CL} proposed an iterative solution based on the Arimoto-Blahut algorithm which achieves excellent performance. However, the method suffers from slow convergence when the number of reflecting elements $N$ is large and the computational complexity of optimizing the reflect beamforming is $\mathcal{O}(N^3)$ per iteration.

{In this work, we propose to directly solve the sum rate maximization problem for a MISO system instead of applying the Arimoto-Blahut algorithm, which is challenging due to the complicated objective function even for the MISO case.} Concretely, in each iteration of the proposed algorithm, a semi-closed form solution to each transmit beamformer is derived. On the other hand, given both transmit beamformers, we derive a closed-form solution to the reflect beamformer. {Compared to the method in \cite{Zhang2020CL}, the proposed algorithm has much faster convergence speed  and the computational complexity of reflect beamforming optimization per iteration is drastically reduced to $\mathcal{O}(N^2)$ without compromising performance. Compared to \cite{Xu2020Arxiv} where the semidefinite relaxation (SDR) technique was used for the reflect beamforming optimization, we obtain a closed-form solution to the reflect beamforming in each iteration which requires much lower computational complexity.}

\emph{Notations:} Vectors and matrices are represented by boldface lower-case and boldface upper-case letters, respectively. $(\cdot)^*$, $(\cdot)^{T}$, $(\cdot)^{H}$, and $\otimes$  denote the conjugate, the transpose, the Hermitian, and the  Kronecker product, respectively. $|a|$ and $\|\mathbf a\|$ are the absolute value of scalar $a$ and the $\ell_2$ norm of  vector $\mathbf a$, respectively. $\Re(\!a\!)$ and $\text{arg}(\!a\!)$ return the real part and the phase of scalar $a$, respectively. $\text{diag}\{\!\mathbf a\!\}$ represents the diagonal matrix with its diagonal elements being the entries of vector $\mathbf a$.  $\mathbf a(\!1\!:\!N\!)$ returns the first $N$ entries of vector $\mathbf a$. $\mathbf A^{-1}$, $\text{vec}(\mathbf A)$, $\text{tr}(\mathbf A)$, and $\lambda_\text{max}(\mathbf A)$ denote the inversion, the vectorization, the trace, and the maximum eigenvalue of matrix $\mathbf A$, respectively.

\vspace{-0.4cm}
\section{System Model and Problem Formulation}\label{sec:model}
\vspace{-0.1cm}
\subsection{System Model}
We consider an IRS-aided point-to-point FD communication system. Both nodes $S_1$ and $S_2$ operate under the FD mode with non-negligible loop interference (LI). Each FD node is equipped with $M$ transmit antennas and one receive antenna, and the IRS has $N$ passive reflecting elements.

The transmit signal of node $S_i$ is expressed by
\begin{align}\label{eq:Txsignal}
   \mathbf {\tilde x}_{i}={\mathbf w_{i}}x_{i},\ i=1,2,
\end{align}
where $\mathbf w_{i}$ is the transmit beamformer of node $S_i$ and $x_{i}$ is the transmit symbol of node $S_i$ with normalized power. Define
$\bar i \triangleq 3-i$. Then, the received signal of node $S_i$ is given by
\begin{align}\label{eq:Rxsignal}
   y_{i}=&(\mathbf h_{IS_i}^H\mathbf \Theta\mathbf H_{S_{\bar i}I}+\mathbf h_{S_{\bar i}S_i}^H)\mathbf {\tilde x}_{{\bar i}}+\mathbf h_{IS_i}^H\mathbf \Theta\mathbf H_{S_iI}\mathbf {\tilde x}_{i}\nonumber \\&+\mathbf h_{S_iS_i}^H \mathbf {\tilde x}_{{i}}+z_{i},\ i=1,2,
\end{align}
where the above four terms represent the information-bearing signal transmitted from node $S_{\bar i}$, the self-interference (SI) transmitted from node $S_i$, the LI due to the FD mechanism of node $S_i$,  and the AWGN at node $S_i$ with variance $\sigma_i^2$, respectively. $\mathbf \Theta\triangleq\text{diag}\{[e^{j\psi_1},\cdots,e^{j\psi_N}]\}$ stands for the IRS reflect beamforming where $\psi_n,\ n=1,\cdots,N$ is the phase shift incurred by the $n$-th reflecting element. $\mathbf h_{IS_i}^H$, $\mathbf H_{S_{\bar i}I}$, $\mathbf h_{S_{\bar i}S_i}^H$, $\mathbf H_{S_{i}I}$, and $\mathbf h_{S_iS_i}^H$ denote the channel from the IRS to node $S_i$, the channel from node $S_{\bar i}$ to the IRS, the channel from  node $S_{\bar i}$ to node $S_i$, the channel from node $S_i$ to the IRS, and the LI channel of node $S_i$, respectively. {Since the path loss of $\mathbf h_{IS_i}^H$ and $\mathbf H_{S_iI}$ is much larger than that of $\mathbf h_{S_iS_i}^H$, the reflecting SI is much weaker than the LI. Hence, we neglect $\mathbf h_{IS_i}^H\mathbf \Theta\mathbf H_{S_iI}\mathbf {\tilde x}_{i}$ as in \cite{Zhang2020CL,Xu2020Arxiv} and update \eqref{eq:Rxsignal} by}
\begin{align}\label{eq:Rxsignal1}
   \tilde y_{i}=&(\mathbf h_{IS_i}^H\mathbf \Theta\mathbf H_{S_{\bar i}I}+\mathbf h_{S_{\bar i}S_i}^H)\mathbf {\tilde x}_{{\bar i}}+\mathbf h_{S_iS_i}^H \mathbf {\tilde x}_{{i}}+z_{i},\ i=1,2.
\end{align}

\vspace{-0.65cm}
\subsection{Sum Rate Maximization Problem}
We aim to maximize the system sum rate by jointly optimizing the IRS reflect beamformer and the transmit beamformers of both FD nodes. According to \eqref{eq:Txsignal} and \eqref{eq:Rxsignal1}, the achievable rate of the link from node $S_{\bar i}$ to node $S_{i}$  equals $R_{i}(\mathbf{w}_{i},\mathbf{w}_{{\bar i}},\mathbf{\Theta})\!\!=\!\! \log_2(\!1\!+\!{|(\mathbf h_{IS_i}^H\mathbf \Theta\mathbf H_{S_{\bar i}I}\!+\!\mathbf h_{S_{\bar i}S_i}^H)\mathbf w_{{\bar i}}|^2}/{(|\mathbf h_{S_iS_i}^H \mathbf w_{{i}}|^2\!+\!\sigma_{i}^2)}\!)$.
Furthermore, we impose a power constraint on $\mathbf w_{i}$ and unit modulus constraints on the diagonal elements of $\mathbf{\Theta}$. Accordingly, we formulate the problem of interest as
\begin{align}\label{eq:SumRateMaxProb}
\mathop{\text{maximize}}\limits_{\mathbf{w}_{i},\mathbf{w}_{\bar i},\mathbf{\Theta}} \quad &
\sum_{i=1}^2R_{i}(\mathbf{w}_{i},\mathbf{w}_{{\bar i}},\mathbf{\Theta})
\nonumber \\ \text{subject to}\quad & \|\mathbf w_{i}\|^2  \leq P_i,\ i=1,2,\nonumber \\& |\theta_n|=1,\ n=1,\cdots,N,
\end{align}
where $\theta_n$ is the $n$-th diagonal of $\mathbf{\Theta}$. This problem cannot be readily solved due to the non-concave objective function and the difficult unit modulus constraints.
\vspace{-0.4cm}
\section{Joint Transmit and Reflect Beamforming Optimization for IRS-Aided FD System}\label{sec:SecRateSingle}
To deal with problem \eqref{eq:SumRateMaxProb}, we first optimize each transmit beamformer by fixing other two variables, which yields a semi-closed form solution. Then, with both $\mathbf w_{1}$ and $\mathbf w_{2}$ fixed, we successfully acquire a closed-form solution to $\mathbf{\Theta}$.
\vspace{-0.3cm}
\subsection{Optimization of $\mathbf w_{i}$ With Given $\mathbf w_{{\bar i}}$ and $\mathbf{\Theta}$}\label{subsec:wi}
Since the problems with respect to $\mathbf w_{i}$ and $\mathbf w_{\bar i}$ are similar, we only focus on the former one without loss of generality.

When $\mathbf w_{{\bar i}}$ and $\mathbf{\Theta}$ are fixed, problem \eqref{eq:SumRateMaxProb} can be recast by
\begin{align}\label{eq:SumRateMaxProbwi}
\mathop{\text{maximize}}\limits_{\mathbf{w}_i} \quad & \frac{c_{\bar i}}{|\mathbf h_{S_{i}S_{i}}^H \mathbf w_{{i}}|^2\!+\!\sigma_{i}^2}\!\!+\! \frac{|\mathbf{h}_{i}^H \mathbf w_i|^2}{\tilde c_{\bar i}}\!+\!\frac{{c_{\bar i}}|\mathbf{h}_{i}^H \mathbf w_{ i}|^2}{\tilde c_{\bar i}(|\mathbf h_{S_{i}S_{i}}^H \mathbf w_{{i}}|^2\!+\!\sigma_{i}^2)}  \nonumber \\
\text{subject to}\quad & \|\mathbf w_{i}\|^2  \leq P_i,
\end{align}
where we removed the logarithm operators and the constant term 1, $\mathbf{h}_i \triangleq \mathbf H_{S_{i}I}^H\mathbf \Theta^H\mathbf h_{IS_{\bar i}}+\mathbf h_{S_{i}S_{\bar i}}$, $c_{\bar i}\triangleq|\mathbf{h}_{\bar i}^H \mathbf w_{\bar i}|^2$, and $\tilde c_{\bar i}\triangleq|\mathbf h_{S_{\bar i}S_{\bar i}}^H \mathbf w_{{\bar i}}|^2+\sigma_{\bar i}^2$.

The above problem is still non-convex since the objective function  (denoted by $f(\mathbf w_{i})$) is not concave. To handle this, we resort to maximizing a concave lower bound of the original objective function as shown in the subsequent proposition.
\begin{proposition}\label{prop:fw}
The objective function of problem \eqref{eq:SumRateMaxProbwi} is lower bounded by the following concave function:
\begin{align}\label{eq:fwlb}
f(\mathbf w_{i}) \geq  -\alpha|\mathbf h_{S_{i}S_{i}}^H \mathbf w_{{i}}|^2+2\Re\{\boldsymbol \beta^H \mathbf w_i\}+\gamma,
\end{align}
where $\alpha\triangleq\frac{c_{\bar i}(|\mathbf{h}_{i}^H\mathbf {\tilde w}_{i}|^2+\tilde c_{\bar i})}{{\tilde c}_{\bar i}(|\mathbf h_{S_{i}S_{i}}^H \mathbf {\tilde w}_{{i}}|^2+\sigma_{i}^2)^2}$, $\boldsymbol \beta\triangleq \frac{1}{{\tilde c}_{\bar i}}\left(1+\frac{c_{\bar i}}{|\mathbf h_{S_{i}S_{i}}^H \mathbf {\tilde w}_{{i}}|^2+\sigma_{i}^2}\right)\mathbf{h}_{i}\mathbf{h}_{i}^H\mathbf {\tilde w}_{i}$, $\gamma\triangleq\alpha|\mathbf h_{S_{i}S_{i}}^H \mathbf {\tilde w}_{{i}}|^2+\frac{c_{\bar i}}{|\mathbf h_{S_{i}S_{i}}^H \mathbf {\tilde w}_{{i}}|^2+\sigma_{i}^2}\!- \!\frac{|\mathbf{h}_{i}^H \mathbf {\tilde w}_i|^2}{\tilde c_{\bar i}}\!-\!\frac{{c_{\bar i}}|\mathbf{h}_{i}^H \mathbf {\tilde w}_{ i}|^2}{\tilde c_{\bar i}(|\mathbf h_{S_{i}S_{i}}^H \mathbf {\tilde w}_{{i}}|^2+\sigma_{i}^2)}$, and $\mathbf {\tilde w}_i$ is a given feasible point. The lower bound is achieved when $\mathbf {w}_{i}=\mathbf {\tilde w}_{i}$.
\end{proposition}
\begin{IEEEproof}
See Appendix~\ref{app:prop1}.
\end{IEEEproof}

We adopt the lower bound in \eqref{eq:fwlb} as a surrogate objective function of problem \eqref{eq:SumRateMaxProbwi}. Accordingly, the optimization problem with respect to $\mathbf{w}_i$ is updated by the following convex quadratically constrained quadratic program (QCQP):
\begin{align}\label{eq:SumRateMaxProbwiQCQP}
\mathop{\text{maximize}}\limits_{\mathbf{w}_i} \quad & -\alpha|\mathbf h_{S_{i}S_{i}}^H \mathbf w_{{i}}|^2+2\Re\{\boldsymbol \beta^H \mathbf w_i\} \nonumber \\
\text{subject to}\quad & \|\mathbf w_{i}\|^2  \leq P_i.
\end{align}
Similarly to \cite[Section III-C]{Zhang2020CL}, we obtain a semi-closed form optimal solution to the above problem by
\begin{align}\label{eq:wopt}
\mathbf{w}_i^\star=(\alpha\mathbf h_{S_{i}S_{i}}\mathbf h_{S_{i}S_{i}}^H+\nu^\star\mathbf I)^{-1}\boldsymbol \beta,
\end{align}
where $\nu^\star$ is the optimal dual variable associated with the power constraint. {It can be readily shown that $\nu^\star$ can be efficiently found by performing a bisection search over the interval $\left[0,{\|\boldsymbol \beta\|}/{\sqrt{P_i}}\right]$.}

%

\vspace{-0.5cm}
\subsection{Optimization of $\mathbf \Theta$ With Given $\mathbf w_1$ and $\mathbf w_{2}$}\label{subsec:theta}
We now investigate the more challenging subproblem with respect to $\mathbf \Theta$ with $\mathbf w_1$ and $\mathbf w_{2}$ fixed, which is expressed by
\begin{align}\label{eq:SumRateMaxProbtheta}
\mathop{\text{maximize}}\limits_{\mathbf{\Theta}} \quad & |\mathbf h_{IS_1}^H\mathbf \Theta\mathbf {\tilde h}_{S_{2}I}+{\tilde h}_{S_{2}S_1}|^2 + |\mathbf h_{IS_2}^H\mathbf \Theta\mathbf {\tilde h}_{S_{1}I}+{\tilde h}_{S_{1}S_2}|^2\nonumber\\& + |\mathbf h_{IS_1}^H\mathbf \Theta\mathbf {\tilde h}_{S_{2}I}+{\tilde h}_{S_{2}S_1}|^2|\mathbf h_{IS_2}^H\mathbf \Theta\mathbf {\tilde h}_{S_{1}I}+{\tilde h}_{S_{1}S_2}|^2\nonumber\\
\text{subject to}\quad & |\theta_n|=1,\ n=1,\cdots,N,
\end{align}
where $\mathbf {\tilde h}_{S_{\bar i}I}\!\!\triangleq\!\!\frac{\mathbf {H}_{S_{\bar i}I}\mathbf w_{\bar i}}{\sqrt{|\mathbf h_{S_iS_i}^H \mathbf w_{{i}}|^2+\sigma_{i}^2}}$ and ${\tilde h}_{S_{\bar i}S_i}\!\!\triangleq\!\!\frac{\mathbf h_{S_{\bar i}S_i}^H\mathbf w_{{\bar i}}}{\sqrt{|\mathbf h_{S_iS_i}^H \mathbf w_{{i}}|^2+\sigma_{i}^2}},\ i=1,2$.
To simplify the objective function, we define $\boldsymbol { \theta}\triangleq[\theta_1^*,\cdots,\theta_{N}^*]^T$ and rewrite $\mathbf h_{IS_i}^H\mathbf \Theta\mathbf {\tilde h}_{S_{\bar i}I}$ by $\boldsymbol { \theta}^H \text{diag}\{\mathbf h_{IS_i}^H\}\mathbf {\tilde h}_{S_{\bar i}I}$. By further introducing a slack variable $t$ with unit norm and defining $\boldsymbol { \bar \theta}\triangleq[\boldsymbol { \theta}^T\ t]^T$, we convert problem \eqref{eq:SumRateMaxProbtheta} to
\begin{align}\label{eq:SumRateMaxProbtheta1}
\mathop{\text{maximize}}\limits_{\boldsymbol{\bar \theta}}\quad & g(\boldsymbol{\bar \theta}) \triangleq |\boldsymbol{\bar \theta}^H\boldsymbol \phi_1|^2+|\boldsymbol{\bar \theta}^H\boldsymbol \phi_2|^2+|\boldsymbol{\bar \theta}^H\boldsymbol \phi_1|^2|\boldsymbol{\bar \theta}^H\boldsymbol \phi_2|^2 \nonumber\\
\text{subject to}\quad & |\bar \theta_n|=1,\ n=1,\cdots,N+1,
\end{align}
where $\boldsymbol \phi_{i}\triangleq[(\text{diag}\{\mathbf h_{IS_i}^H\}\mathbf {\tilde h}_{S_{\bar i}I})^T\ {\tilde h}_{S_{\bar i}S_i}]^T,\ i=1,2$. For this problem, even if we can, analogously to solving problem \eqref{eq:SumRateMaxProbwi}, determine a concave quadratic lower bound to the objective function, the resultant problem is still non-convex whose optimal solution cannot be readily obtained. To handle this, we derive an affine lower bound as a surrogate objective function, which further enables us to achieve a closed-form solution.
\begin{proposition}\label{prop:ftheta}
The objective function of problem \eqref{eq:SumRateMaxProbtheta1} is lower bounded by the following affine function:
\begin{align}\label{eq:gthetalb}
g(\boldsymbol{\bar \theta}) \geq  & \Re\{\boldsymbol \rho^H\boldsymbol{\bar \theta}\}+\kappa,
\end{align}
where $\boldsymbol \rho \triangleq 2(\sum_{i=1}^2  \boldsymbol \phi_{i} \boldsymbol \phi_{i}^H+\lambda_\text{max}(\mathbf \Psi)\mathbf I-\mathbf \Psi)\boldsymbol{\tilde \theta}$,  $\kappa\triangleq-2(N+1)\lambda_\text{max}(\mathbf \Psi)-|\boldsymbol{\tilde \theta}^H\boldsymbol \phi_1|^2-|\boldsymbol{\tilde \theta}^H\boldsymbol \phi_2|^2-3|\boldsymbol{\tilde \theta}^H\boldsymbol \phi_1|^2|\boldsymbol{\tilde \theta}^H\boldsymbol \phi_2|^2$, $\mathbf \Psi \triangleq -(\boldsymbol \phi_2\boldsymbol \phi_2^H\boldsymbol{\tilde \theta}\boldsymbol{\tilde \theta}^H \boldsymbol \phi_1\boldsymbol \phi_1^H+\boldsymbol \phi_1\boldsymbol \phi_1^H\boldsymbol{\tilde \theta}\boldsymbol{\tilde \theta}^H \boldsymbol \phi_2\boldsymbol \phi_2^H)$, and $\boldsymbol{\tilde \theta}$ is a given feasible point. The lower bound is achieved when $\boldsymbol{\bar \theta}=\boldsymbol{\tilde \theta}$.
\end{proposition}
\begin{IEEEproof}
See Appendix~\ref{app:prop2}.
\end{IEEEproof}

By replacing the objective function of problem \eqref{eq:SumRateMaxProbtheta1} with the lower bound provided in \eqref{eq:gthetalb}, we attain
\begin{align}\label{eq:SumRateMaxProbtheta2}
\mathop{\text{maximize}}\limits_{\boldsymbol{\bar \theta}}\quad & \Re\{\boldsymbol \rho^H\boldsymbol{\bar \theta}\} \nonumber\\
\text{subject to}\quad & |\bar \theta_n|=1,\ n=1,\cdots,N+1.
\end{align}
The optimal solution to this problem is given by
\begin{align}\label{eq:thetabaropt}
\bar \theta_n^\star=e^{j\text{arg}\{\rho_n\}},\ n=1,\cdots,N+1,
\end{align}
{where $\rho_n$ is the $n$-th entry of $\boldsymbol \rho$.} Moreover, according to the definitions of $\boldsymbol{\bar \theta}$ and $\boldsymbol{\theta}$, the solution to problem \eqref{eq:SumRateMaxProbtheta} is
\begin{align}\label{eq:Thetaopt}
\mathbf \Theta^\star=\text{diag}\{(\boldsymbol{\bar \theta}^\star(1:N)/\bar \theta_{N+1}^\star)^*\}.
\end{align}
{Note that $\mathbf w_i^\star$ and $\mathbf \Theta^\star$ are not necessarily optimal solutions. However, based on the two solutions, we can still develop a convergent algorithm for problem \eqref{eq:SumRateMaxProb} in the next subsection.}
\vspace{-0.5cm}
\subsection{Alternating Algorithm for Problem \eqref{eq:SumRateMaxProb}}
The proposed algorithm for problem \eqref{eq:SumRateMaxProb} is summarized in Algorithm~\ref{alg1} whose convergence is proved as follows.
\begin{proposition}\label{prop:conv}
Algorithm~\ref{alg1} yields a convergent solution.
\end{proposition}
\begin{IEEEproof}
See Appendix~\ref{app:prop3}.
\end{IEEEproof}
\begin{remark}
It can be analyzed that the computational complexity per iteration of  Algorithm~\ref{alg1} is $\mathcal{O}(M^3+N^2)$\footnote{For the optimization of $\mathbf \Theta$, we can use the power method with deflation \cite{Wilkinson1965Book} to perform the eigenvalue decomposition of $\mathbf \Phi$. The corresponding complexity is $\mathcal{O}(N^2)$ since $\mathbf \Phi$ is a rank-2 matrix.}. Moreover, the convergence of Algorithm~\ref{alg1} can be further accelerated by applying the acceleration scheme based on SQUAREM \cite[Section V-B]{Song2016TSP} for the optimization of $\mathbf \Theta$ (see Table~\ref{table:iteration}).
\end{remark}
\begin{remark}
The main differences between Algorithm~\ref{alg1} and the method in \cite{Zhang2020CL} are twofold: 1) {we address the sum rate maximization problem straightforwardly instead of applying the Arimoto-Blahut structure to convert the original problem to a new form with two more auxiliary variables, which may account for its faster convergence}; 2) In each iteration, we obtain a closed-form solution to $\mathbf \Theta$ with computational complexity $\mathcal{O}(N^2)$ while the solution to $\mathbf \Theta$ in \cite{Zhang2020CL} has an order-of-magnitude higher computational complexity $\mathcal{O}(N^3)$.
\end{remark}
\begin{remark}
{Since $\mathbf h_i^H=(\boldsymbol \theta')^H \mathbf {\bar H}_{S_iS_{\bar i}}$ with $\boldsymbol \theta' \triangleq [\theta_1^*,\cdots,\theta_N^*,1]$ and $\mathbf {\bar H}_{S_iS_{\bar i}} \triangleq [ \mathbf H_{S_iI}^H \text{diag}\{\mathbf h_{IS_{\bar i}} \} \  \mathbf h_{S_iS_{\bar i}}]^H$, it suffices to know $\mathbf {\bar H}_{S_iS_{\bar i}}$ for the proposed algorithm, which can be estimated with the scheme developed in \cite{You2019Arxiv}.}
\end{remark}
\begin{remark}
{For the phase shift constraint $|\theta_n| \leq 1,\ n=1,\cdots,N$, it can be shown that we only need to update \eqref{eq:thetabaropt} by $\bar \theta_n^\star\!\!=\!\!\min\left\{{|\rho_n|}/{(2\lambda_\text{max}(\mathbf \Psi))},1\right\}e^{j\text{arg}\{\rho_{n}\}},\ n=1,\!\cdots\!,N$ while with $\bar \theta_{N+1}^\star$ unchanged. For the discrete phase shift constraint, we can acquire a high-quality solution using Algorithm~\ref{alg1} and the quantization based technique in \cite[Section III-D]{Zhang2020CL}.}
\end{remark}

\begin{algorithm}[t]
\caption{\textbf{Proposed algorithm for problem \eqref{eq:SumRateMaxProb}}}\label{alg1}

\begin{algorithmic}[1]

\STATE \textit{Initialization:}  set initial $\mathbf {\tilde w}_{1}$, $\mathbf {\tilde w}_{2}$, $\boldsymbol {\tilde \Theta}$, and convergence accuracy $\epsilon$.
\STATE \textbf{repeat}

\STATE \text{  } Fix $\mathbf {w}_{2}=\mathbf {\tilde w}_{2}$ and $\boldsymbol \Theta=\boldsymbol {\tilde \Theta}$, and obtain $\mathbf {w}_{1}^\star$ using \eqref{eq:wopt}.

\STATE \text{  } Fix $\mathbf {w}_{1}=\mathbf {w}_{1}^\star$ and $\boldsymbol \Theta=\boldsymbol {\tilde \Theta}$, and obtain $\mathbf {w}_{2}^\star$ using \eqref{eq:wopt}.

\STATE \text{  } Fix $\mathbf {w}_{1}=\mathbf {w}_{1}^\star$ and $\mathbf {w}_{2}=\mathbf {w}_{2}^\star$,  and calculate $\mathbf \Theta^\star$ using \eqref{eq:thetabaropt} and \eqref{eq:Thetaopt}.

\STATE \text{  } Set $\mathbf {\tilde w}_{i}=\mathbf {w}_{i}^\star,\ i=1,2$ and $\boldsymbol {\tilde \Theta}=\mathbf \Theta^\star$.

\STATE \textbf{until} convergence.

\STATE Output $\mathbf {w}_{1}^\star$, $\mathbf {w}_{2}^\star$, and $\mathbf \Theta^\star$.
\end{algorithmic}
\end{algorithm}
\vspace{-0.5cm}
\section{Simulation Results}\label{sec:simu}
\begin{figure}[t]
\begin{minipage}[t]{0.5\linewidth}
\centering
\includegraphics[width=1.6in]{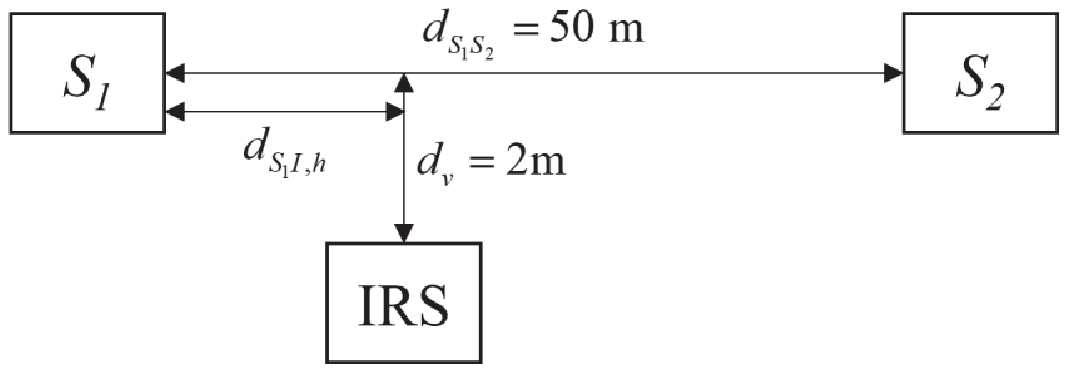}
\caption{{Simulation setup.}}\label{fig:geometry}
\end{minipage}%
\begin{minipage}[t]{0.5\linewidth}
\centering
\includegraphics[width=1.5in]{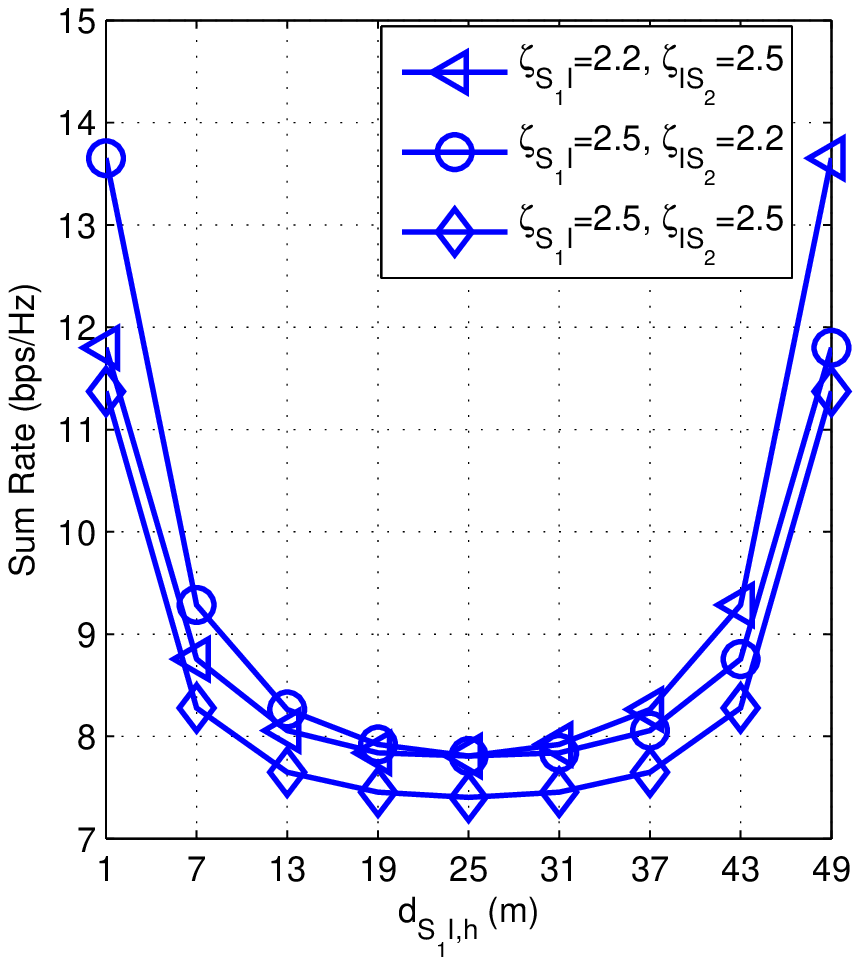}
\caption{{Sum rate versus $d_{S_1I,h}$.}}\label{fig:darh}
\end{minipage}
\end{figure}

We conduct simulations to test the proposed algorithm.
We set $M=4$, $N=40$, $P_1=P_2=15\ \text{dBW}$, and $\sigma_{1}^2=\sigma_{2}^2=-80\ \text{dBW}$.
The path loss of both LI channels is $-90$ dB due to the LI cancellation. For other channels, the path loss at distance $d$ is given by $\xi=( \tilde \xi-10\zeta\log_{10}({d}/{\tilde d}))\ \text{dB}$,
where $\tilde \xi$ is the path loss at the reference distance $\tilde d$, and $\zeta$ denotes the path loss exponent (PLE). We set $\tilde \xi=-30\text{ dB}$ and $\tilde d=1 \text{m}$. The PLE of the channel $\mathbf H_{S_iI}$ and $\mathbf h_{I S_{\bar i}}^H,\ i=1,2$ are set to $\zeta_{S_iI}=\zeta_{IS_{\bar i}}=2.5$, and the PLE of the channel $\mathbf h_{S_iS_{\bar i}}^H$ is set to $\zeta_{S_iS_{\bar i}}=3.5$. The distance of all links is calculated according to Fig.~\ref{fig:geometry}, where the IRS lies in a horizontal line that is parallel to the one between node $S_1$ and node $S_2$.  {We adopt the Rician model for the LI channel with the Rician factor being 5 dB \cite{Sun2017TCOM,Aslani2019TVT} and use the Rayleigh model for other channels.}

%

\begin{figure}[t]
\begin{minipage}[t]{0.5\linewidth}
\centering
\includegraphics[width=1.5in]{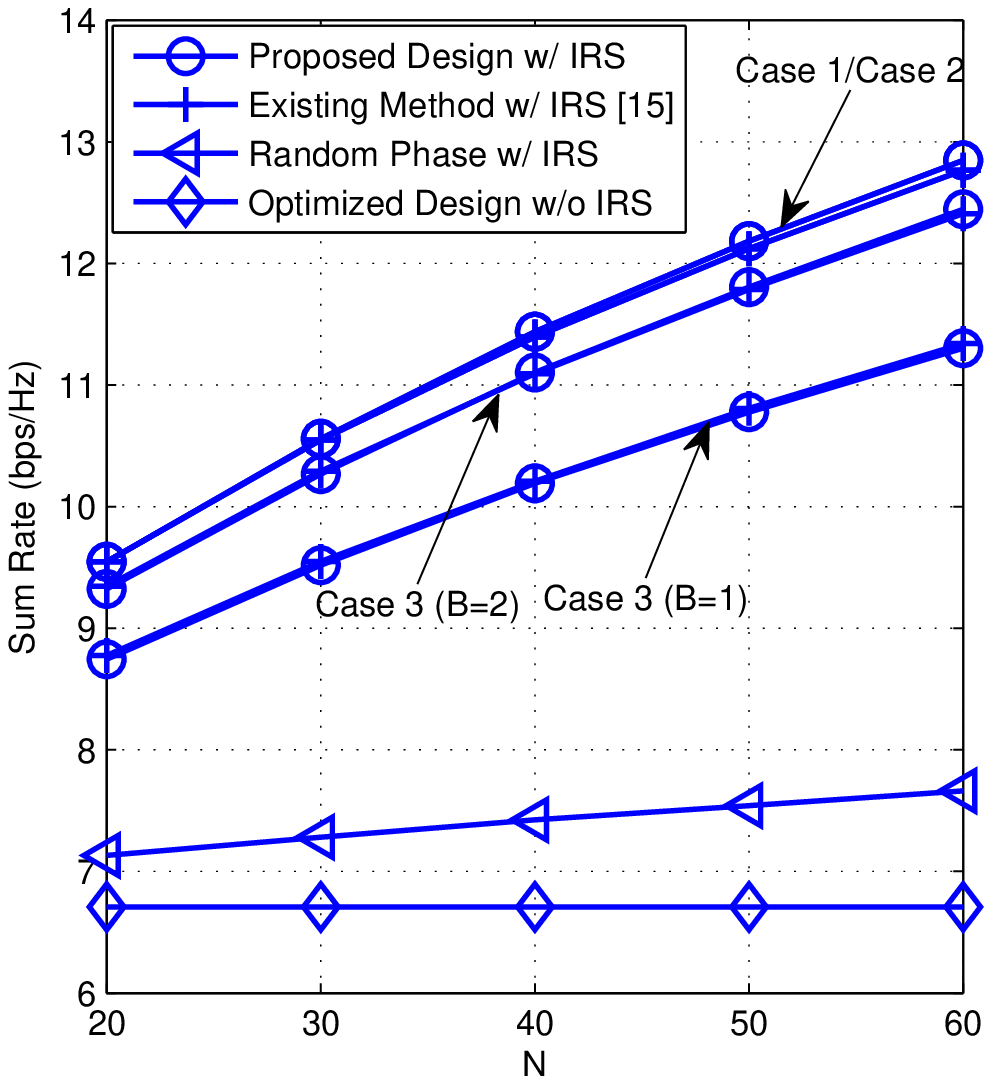}
\caption{{Sum rate versus $N$.}}\label{fig:RateN}
\end{minipage}%
\begin{minipage}[t]{0.5\linewidth}
\centering
\includegraphics[width=1.5in]{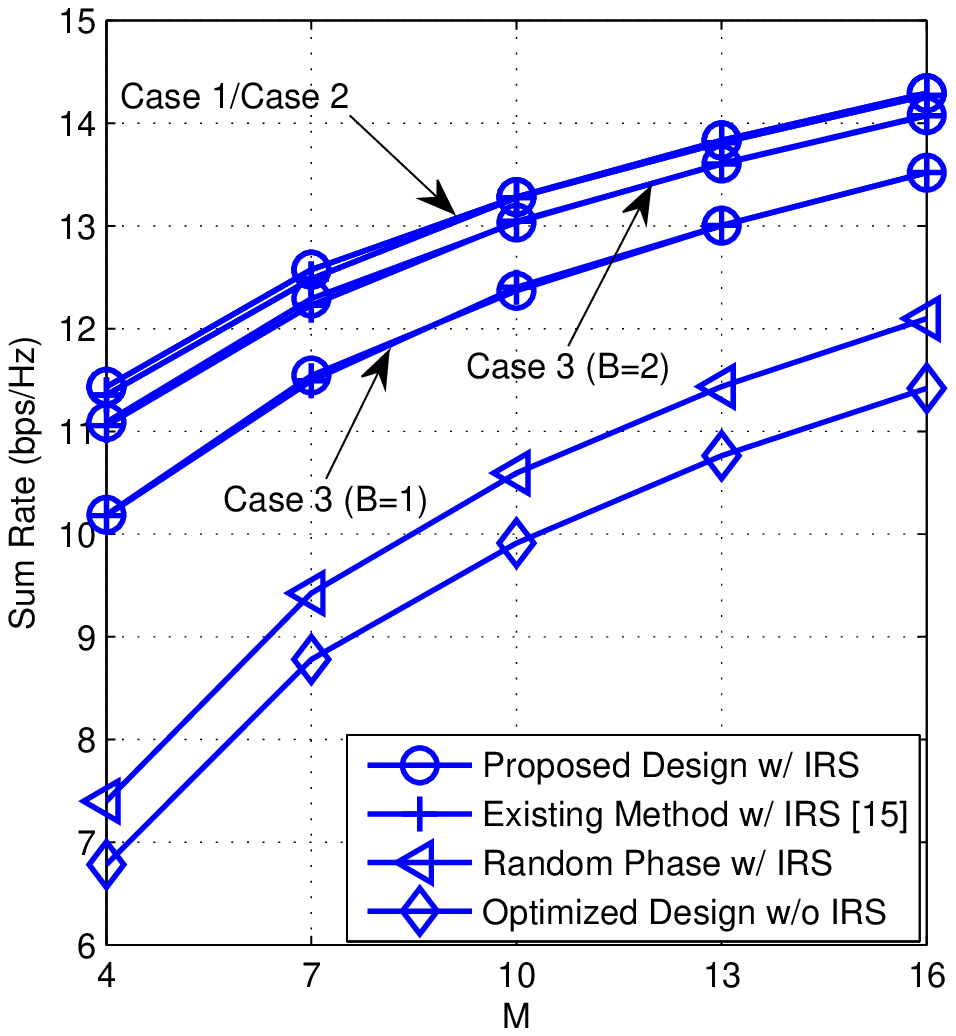}
\caption{{Sum rate versus $M$.}}\label{fig:RateM}
\end{minipage}
\end{figure}

{We show the sum rate performance versus $d_{S_1I,h}$ in Fig.~\ref{fig:darh}. We can observe that the sum rate gradually increases when the IRS gets close to either node $S_1$ or node $S_2$ since the reflect beamforming gain becomes larger. In particular, when $\zeta_{S_1I}=\zeta_{IS_2}$, the sum rate curve is symmetric with respect to the midpoint $d_{S_1I,h}=25\ \text{m}$. This is because, the path losses of the reflected links corresponding to any two symmetric points  are the same. On the other hand, when $\zeta_{S_1I} \neq \zeta_{IS_2}$, the sum rate curve is asymmetric and a higher sum rate can be achieved when the IRS approaches node $S_i$ where $i$ satisfies $\zeta_{S_iI} > \zeta_{IS_{\bar i}}$. This is because, given the same distance, the channel between node $S_i$ and the IRS is subject to severer path loss than the channel between node $S_{\bar i}$ and the IRS.}


{In Fig.~\ref{fig:RateN}, we compare the proposed method with three benchmark schemes: 1) existing solution based on the Arimoto-Blahut algorithm \cite{Zhang2020CL}; 2) random IRS phase shift design; 3) optimized beamforming design for the FD system without IRS. The third scheme is achieved by setting $\mathbf \Theta$ to zero in Algorithm~\ref{alg1}. Cases 1, 2, and 3 refer to the constraints $|\theta_n| \leq 1$, $|\theta_n| = 1$, and $\theta_n \in \left\{0,{2\pi}/{2^B},\cdots,{2\pi(2^B-1)}/{2^B}\right\}$, respectively, where $B$ denotes the number of bits used to represent the phase shift levels. It can be found that the use of IRS can significantly enhance the sum rate especially for large $N$, which is due to the reflect beamforming gain provided by the IRS. Compared to the random phase shift scheme, the proposed method achieves much higher rate since we optimize the phase shifts of IRS.
For the proposed design and the existing method in \cite{Zhang2020CL}, the sum rates under Case 1 and Case 2 coincide and the rate gap between Case 1/Case 2 and Case 3 (B=2) is small, which are consistent with the results in \cite{Zhang2020CL}. Moreover, the proposed design achieves almost the same performance as the existing method in \cite{Zhang2020CL} under all 3 cases because we also aim at maximizing the sum rate. We show the sum rate versus the number of antennas $M$ in Fig.~\ref{fig:RateM}, where we observe similar phenomenon as in Fig.~\ref{fig:RateN}. Besides, the gain due to the use of IRS or the optimization of IRS phase shifts is especially evident for relatively small $M$ since the transmit beamforming gain becomes more dominant for large $M$.}

\begin{table*}[t]
\renewcommand{\arraystretch}{1.3}
\caption{{Comparison of Average Number of Iterations (Convergence Accuracy $\epsilon=10^{-3}$)}}
\label{table:iteration}
\centering
\begin{tabular}{|c|c|c|c|c|c|c|c|c|c|}
\hline \multirow{3}*{Design Method} & \multicolumn{9}{c|}{Average Number of Iterations}\\ \cline{2-10} & \multicolumn{3}{c|}{$d_{S_1S_2}=50\ \text{m}$,\ $P=15\ \text{dBW}$} & \multicolumn{3}{c|}{$d_{S_1S_2}=45\ \text{m}$,\ $P=15\ \text{dBW}$} & \multicolumn{3}{c|}{$d_{S_1S_2}=50\ \text{m}$,\ $P=12\ \text{dBW}$} \\
\cline{2-10}
& $N=20$ & $N=40$ & $N=60$ & $N=20$ & $N=40$ & $N=60$ & $N=20$ & $N=40$ & $N=60$\\
\hline
Method in [15] & $27.9933 $ &   $ 49.2833$  &   $59.55$ &   $12.13 $ &   $18.99$  &   $26.1667$ & $19.3767 $ &   $35.75$  &   $45.68$\\
\hline
Proposed w/o Acceleration & $13.3533$  & $ 21.9533$   & $28.5933$ & $5.5233$  & $ 9.1367$   & $12.6767$ & $12.2467$  & $ 20.5567$   & $27.57$\\
\hline
Proposed w/ Acceleration & $7.5767$   & $10.82$  & $ 13.9033$ & $4.1767$   & $5.5667$  & $ 6.8667$ & $7.25$   & $11.1233$  & $13.9933$\\
\hline
\end{tabular}
\end{table*}
{As shown in Table~\ref{table:iteration},  the proposed method without acceleration requires much fewer iterations to reach convergence than the method in \cite{Zhang2020CL}. Moreover, the average number of iterations can be further reduced after we apply the acceleration scheme.}

\section{Conclusions}
We studied the sum rate maximization for an IRS-aided FD system by jointly optimizing the transmit beamforming and the IRS reflect beamforming. To address the difficult non-convex problem, we developed a fast converging iterative algorithm where the transmit beamformer and the reflect beamformer admit a semi-closed form solution and a closed-form solution, respectively, in each iteration. Compared to an existing scheme based on the Arimoto-Blahut algorithm, the proposed method has clear superiority in terms of convergence speed and computational complexity. {Future works include convergence speed analysis and the extensions to the MIMO scenario and the robust beamforming design.}

\begin{appendices}
\vspace{-0.4cm}
\section{Proof of \emph{Proposition}~\ref{prop:fw}}\label{app:prop1}
To simply the notation, let us define $f_1(\mathbf w_{i})\triangleq{|\mathbf{h}_{i}^H \mathbf w_{ i}|^2}$, $f_2(\mathbf w_{i}) \triangleq \frac{1}{|\mathbf h_{S_{i}S_{i}}^H \mathbf w_{{i}}|^2+\sigma_{i}^2}$, and $f_3(\mathbf w_{i})\triangleq\frac{|\mathbf{h}_{i}^H \mathbf w_{ i}|^2}{|\mathbf h_{S_{i}S_{i}}^H \mathbf w_{{i}}|^2+\sigma_{i}^2}$.

Since $f_1(\mathbf w_{i})$ is convex with respect to $\mathbf w_i$, it is lower bounded by its first-order Taylor expansion at given $\mathbf {\tilde w}_{i}$, i.e.,
\begin{align}
f_1(\mathbf w_{i}) \geq &f_1(\mathbf {\tilde w}_{i})+ 2\Re\{\mathbf {\tilde w}_{i}^H \mathbf{h}_{i}\mathbf{h}_{i}^H(\mathbf w_{ i}-\mathbf {\tilde w}_{i})\}.\label{eq:f1}
\end{align}

For $f_2(\mathbf w_{i})$, we first rewrite it by $f_2(u)={1}/{u}$, where $u \triangleq |\mathbf h_{S_{i}S_{i}}^H \mathbf w_{{i}}|^2+\sigma_{i}^2$. Clearly, $f_2(u)$ is a convex function and is thus lower bounded by $f_2(u) \geq f_2(\tilde u)-{(u-\tilde u)}/{\tilde u^2}$, where $\tilde u \triangleq |\mathbf h_{S_{i}S_{i}}^H \mathbf {\tilde w}_{{i}}|^2+\sigma_{i}^2$. Therefore, we further have
\begin{align}
 f_2(\!\mathbf w_{i}\!)\!\!\geq \!\! f_2(\!\mathbf {\tilde w}_{i}\!)\!\!-\!\!{|\!\mathbf h_{S_{i}S_{i}}^H \!\mathbf w_{{i}}\!|^2\!\!-\!\!|\!\mathbf h_{S_{i}S_{i}}^H \!\mathbf {\tilde w}_{{i}}\!|^2}\!/\!{(|\!\mathbf h_{S_{i}S_{i}}^H\! \mathbf {\tilde w}_{{i}}\!|^2\!\!+\!\!\sigma_{i}^2)^2}.\label{eq:f2}
\end{align}

We express $f_3(\mathbf w_{i})$ by $f_3(\mathbf w_{i},u)={|\mathbf{h}_{i}^H \mathbf w_{ i}|^2}/{u}$. Since $f(x,y)={|x|^2}/{y}$ is jointly convex with $(x,y)$ for $y>0$ \cite[Section 3.1.7]{BoydConvex} and $\mathbf{h}_{i}^H\mathbf w_{ i}$ is affine with respect to $\mathbf w_{i}$, $f_3(\mathbf w_{i},u)$ is jointly convex with $(\mathbf w_{i},u)$. Thus, based on the Taylor expansion, it follows that $f_3(\mathbf w_{i},u) \geq f_3(\mathbf {\tilde w}_{i},\tilde u)+{2\Re\left\{\mathbf {\tilde w}_{i}^H\mathbf{h}_{i}\mathbf{h}_{i}^H(\mathbf w_{ i}-\mathbf {\tilde w}_{i})\right\}}/{\tilde u}-{|\mathbf{h}_{i}^H\mathbf {\tilde w}_{i}|^2(u-\tilde u)}/{\tilde u^2}$. Furthermore, using the definitions of $u$ and $\tilde u$, we have
\begin{align}\label{eq:f3}
&f_3(\mathbf w_{i}) \!\!\geq \!\!f_3(\mathbf {\tilde w}_{i})\!\!+\!\!{2\Re\left\{\mathbf {\tilde w}_{i}^H\mathbf{h}_{i}\mathbf{h}_{i}^H(\mathbf w_{ i}\!\!-\!\!\mathbf {\tilde w}_{i})\right\}}/{(|\mathbf h_{S_{i}S_{i}}^H \mathbf {\tilde w}_{{i}}|^2\!\!+\!\!\sigma_{i}^2)}\nonumber \\ & -\!\!{|\mathbf{h}_{i}^H\mathbf {\tilde w}_{i}|^2(|\mathbf h_{S_{i}S_{i}}^H \!\mathbf w_{{i}}|^2\!\!-\!\!|\mathbf h_{S_{i}S_{i}}^H \! \mathbf {\tilde w}_{{i}}|^2)}/{\left(|\mathbf h_{S_{i}S_{i}}^H \! \mathbf {\tilde w}_{{i}}|^2\!\!+\!\!\sigma_{i}^2\right)^2}.
\end{align}
Substituting \eqref{eq:f1}--\eqref{eq:f3} into the objective function of problem \eqref{eq:SumRateMaxProbwi}, we obtain \eqref{eq:fwlb}, which is concave with respect to $\mathbf w_{i}$.
\vspace{-0.4cm}
\section{Proof of \emph{Proposition}~\ref{prop:ftheta}}\label{app:prop2}
Define $g_1(\boldsymbol{\bar \theta}) \triangleq |\boldsymbol{\bar \theta}^H\boldsymbol \phi_1|^2$, $g_2(\boldsymbol{\bar \theta})\triangleq |\boldsymbol{\bar \theta}^H\boldsymbol \phi_2|^2$, and $g_3(\boldsymbol{\bar \theta})\triangleq|\boldsymbol{\bar \theta}^H\boldsymbol \phi_1|^2|\boldsymbol{\bar \theta}^H\boldsymbol \phi_2|^2$. Similarly to \eqref{eq:f1} and \eqref{eq:f2}, we readily obtain a lower bound to $g_i(\boldsymbol{\bar \theta})$ by
\begin{align}
g_i(\boldsymbol{\bar \theta})\geq & |\boldsymbol{\tilde \theta}^H\boldsymbol \phi_i|^2+2\Re\{\boldsymbol{\tilde \theta}^H\boldsymbol \phi_i\boldsymbol \phi_i^H (\boldsymbol{\bar \theta}-\boldsymbol{\tilde \theta})\}\nonumber \\ = & 2\Re\{\boldsymbol{\tilde \theta}^H\boldsymbol \phi_i\boldsymbol \phi_i^H \boldsymbol{\bar \theta}\}-|\boldsymbol{\tilde \theta}^H\boldsymbol \phi_i|^2,\ i=1,2.\label{eq:g1}
\end{align}

Different from $g_1(\boldsymbol{\bar \theta})$ or $g_2(\boldsymbol{\bar \theta})$, it is non-trivial to find an appropriate lower bound to $g_3(\boldsymbol{\bar \theta})$. We first rewrite $g_3(\boldsymbol{\bar \theta})$ by
\begin{align}
g_3(\boldsymbol{\bar \theta})&=\!\boldsymbol{\bar \theta}^H\boldsymbol \phi_1\boldsymbol \phi_1^H\boldsymbol{\bar \theta}\boldsymbol{\bar \theta}^H\boldsymbol \phi_2\boldsymbol \phi_2^H\boldsymbol{\bar \theta}\!\overset{(a)}{=}\!\text{tr}(\boldsymbol \phi_1\boldsymbol \phi_1^H\boldsymbol{\bar \theta}\boldsymbol{\bar \theta}^H\boldsymbol \phi_2\boldsymbol \phi_2^H\boldsymbol{\bar \theta}\boldsymbol{\bar \theta}^H)\nonumber\\& \overset{(b)}{=} \text{vec}^T(\boldsymbol{\bar \theta}\boldsymbol{\bar \theta}^H) ( (\boldsymbol \phi_1\boldsymbol \phi_1^H) \otimes (\boldsymbol \phi_2\boldsymbol \phi_2^H)^T )\text{vec}((\boldsymbol{\bar \theta}\boldsymbol{\bar \theta}^H)^T) \nonumber\\ & \overset{(c)}{=} \text{vec}^H(\boldsymbol{\bar \theta}\boldsymbol{\bar \theta}^H) ( (\boldsymbol \phi_1\boldsymbol \phi_1^H)^* \otimes (\boldsymbol \phi_2\boldsymbol \phi_2^H) )\text{vec}(\boldsymbol{\bar \theta}\boldsymbol{\bar \theta}^H),
\end{align}
where (a) holds because $\text{tr}(\mathbf {A B})=\text{tr}(\mathbf {B A})$,  (b) is due to $\text{tr}(\mathbf {A B C D})=\text{vec}^T(\mathbf D)(\mathbf  A \otimes \mathbf  C^T)\text{vec}(\mathbf B^T)$, and (c) holds since $g_3(\boldsymbol{\bar \theta})$ is real, i.e., $g_3(\boldsymbol{\bar \theta})=g_3^*(\boldsymbol{\bar \theta})$. Define $\boldsymbol{\breve \theta}\triangleq\text{vec}(\boldsymbol{\bar \theta}\boldsymbol{\bar \theta}^H)$, $\boldsymbol{\hat \theta}\triangleq\text{vec}(\boldsymbol{\tilde \theta}\boldsymbol{\tilde \theta}^H)$, and $\mathbf \Phi \triangleq (\boldsymbol \phi_1\boldsymbol \phi_1^H)^* \otimes (\boldsymbol \phi_2\boldsymbol \phi_2^H)$. Then, we have
\begin{align}\label{eq:g3a}
g_3(\boldsymbol{\bar \theta})&=\boldsymbol{\breve \theta}^H  \mathbf \Phi \boldsymbol{\breve \theta} \overset{(a)}{\geq}  \boldsymbol{\hat \theta}^H \mathbf { \Phi} \boldsymbol{\hat \theta}+2\Re\{\boldsymbol{\hat \theta}^H \mathbf {\Phi} (\boldsymbol{\breve \theta}-\boldsymbol{\hat \theta})\}\nonumber \\& \overset{(b)}{=}\boldsymbol{\hat \theta}^T \mathbf {\Phi}^* \boldsymbol{\breve \theta}^*+\boldsymbol{\breve \theta}^T \mathbf {\Phi}^* \boldsymbol{\hat \theta}^*-\boldsymbol{\hat \theta}^T \mathbf {\Phi}^* \boldsymbol{\hat \theta}^*\nonumber \\& \overset{(c)}{=} \text{tr}(\boldsymbol \phi_1\boldsymbol \phi_1^H\boldsymbol{\bar \theta}\boldsymbol{\bar \theta}^H\boldsymbol \phi_2\boldsymbol \phi_2^H\boldsymbol{\tilde \theta}\boldsymbol{\tilde \theta}^H)\!\!+\!\text{tr}(\!\boldsymbol \phi_1\boldsymbol \phi_1^H\boldsymbol{\tilde \theta}\boldsymbol{\tilde \theta}^H\boldsymbol \phi_2\boldsymbol \phi_2^H\boldsymbol{\bar \theta}\boldsymbol{\bar \theta}^H\!)\nonumber \\& \quad\ -\text{tr}(\boldsymbol \phi_1\boldsymbol \phi_1^H\boldsymbol{\tilde \theta}\boldsymbol{\tilde \theta}^H\boldsymbol \phi_2\boldsymbol \phi_2^H\boldsymbol{\tilde \theta}\boldsymbol{\tilde \theta}^H)\nonumber \\& \overset{(d)}{=} \boldsymbol{\bar \theta}^H(\boldsymbol \phi_2\boldsymbol \phi_2^H\boldsymbol{\tilde \theta}\boldsymbol{\tilde \theta}^H \boldsymbol \phi_1\boldsymbol \phi_1^H+\boldsymbol \phi_1\boldsymbol \phi_1^H\boldsymbol{\tilde \theta}\boldsymbol{\tilde \theta}^H \boldsymbol \phi_2\boldsymbol \phi_2^H)\boldsymbol{\bar \theta}\nonumber \\&\quad\ -|\boldsymbol{\tilde \theta}^H\boldsymbol \phi_1|^2|\boldsymbol{\tilde \theta}^H\boldsymbol \phi_2|^2,
\end{align}
where (a) holds due to the convexity of $\boldsymbol{\breve \theta}^H  \mathbf \Phi \boldsymbol{\breve \theta} $, (b) holds since the terms $\boldsymbol{\hat \theta}^H\mathbf {\Phi} \boldsymbol{\breve \theta}+\boldsymbol{\breve \theta}^H \mathbf {\Phi} \boldsymbol{\hat \theta}$ and $\boldsymbol{\hat \theta}^H \mathbf {\Phi}\boldsymbol{\hat \theta}$ are real numbers, (c) is obtained based on the definitions of $\boldsymbol{\hat \theta}$, $\boldsymbol{\tilde \theta}$, and $\mathbf \Phi$, the fact that $\mathbf X^*=\mathbf X^T$ holds for any Hermitian matrix $\mathbf X$, and the equation $\text{tr}(\mathbf {A B C D})=\text{vec}^T(\mathbf D)(\mathbf  A \otimes \mathbf  C^T)\text{vec}(\mathbf B^T)$, and (d) is derived by invoking $\text{tr}(\mathbf {A B})=\text{tr}(\mathbf {B A})$. Define $\mathbf \Psi \triangleq -(\boldsymbol \phi_2\boldsymbol \phi_2^H\boldsymbol{\tilde \theta}\boldsymbol{\tilde \theta}^H \boldsymbol \phi_1\boldsymbol \phi_1^H+\boldsymbol \phi_1\boldsymbol \phi_1^H\boldsymbol{\tilde \theta}\boldsymbol{\tilde \theta}^H \boldsymbol \phi_2\boldsymbol \phi_2^H)$. Then, by utilizing \cite[Section III-C]{Sun2017TSP} and $\|\boldsymbol{\bar \theta}\|^2=\|\boldsymbol{\tilde \theta}\|^2=N+1$, we have
\begin{align}\label{eq:g3b}
\boldsymbol{\bar \theta}^H \!\mathbf \Psi\! \boldsymbol{\bar \theta}\!\!  \leq \!\! 2\Re\{\!\boldsymbol{\bar \theta}^H(\!\mathbf \Psi\!\!-\!\!\lambda_\text{max}(\!\mathbf \Psi\!)\mathbf I\!)\boldsymbol{\tilde \theta}\!\}\!\!+\!\!2(\!N\!\!+\!\!1\!)\lambda_\text{max}(\!\mathbf \Psi\!)\!\!-\!\!\boldsymbol{\tilde \theta}^H\! \mathbf \Psi \!\boldsymbol{\tilde \theta}.
\end{align}
Based on \eqref{eq:g3a} and \eqref{eq:g3b}, we obtain
\begin{align}\label{eq:g3c}
g_3(\boldsymbol{\bar \theta}) \geq& 2\Re\{\boldsymbol{\bar \theta}^H(\lambda_\text{max}(\mathbf \Psi)\mathbf I-\mathbf \Psi)\boldsymbol{\tilde \theta}\}-2(N+1)\lambda_\text{max}(\mathbf \Psi)\nonumber \\&-3|\boldsymbol{\tilde \theta}^H\boldsymbol \phi_1|^2|\boldsymbol{\tilde \theta}^H\boldsymbol \phi_2|^2.
\end{align}
According to \eqref{eq:g1} and \eqref{eq:g3c}, we eventually obtain \eqref{eq:gthetalb}.
\vspace{-0.5cm}
\section{Proof of \emph{Proposition}~\ref{prop:conv}}\label{app:prop3}
Since the objective function of problem \eqref{eq:SumRateMaxProb} must be upper bounded by a finite value, we only need to prove that the objective value of problem \eqref{eq:SumRateMaxProb} (denoted by $R(\mathbf w_1,\mathbf w_2,\mathbf \Theta)$) keeps increasing after each iteration of Algorithm~\ref{alg1}.

Define the lower bound in \emph{Proposition}~\ref{prop:fw} by $f(\mathbf w_i|\mathbf {\tilde w}_i)$. Then, we have $f(\mathbf {\tilde w}_i)\overset{(a)}{=}f(\mathbf {\tilde w}_i|\mathbf {\tilde w}_i)
\overset{(b)}{\leq} f(\mathbf {w}_i^\star|\mathbf {\tilde w}_i)
\overset{(c)}{\leq} f(\mathbf {w}_i^\star)$
where (a) and (c) hold due to Proposition~\ref{prop:fw}, and (b) holds because  $\mathbf {w}_i^\star$ maximizes $f(\mathbf w_i|\mathbf {\tilde w}_i)$. {Since $R(\mathbf w_1,\mathbf w_2,\mathbf \Theta)=\log_2(f(\mathbf w_i)+1)$,  it follows that $R(\mathbf {\tilde w}_1,\mathbf {\tilde w}_2,\mathbf {\tilde \Theta}) \leq R(\mathbf {w}_1^\star,\mathbf {\tilde w}_2,\mathbf {\tilde \Theta})$ and $R(\mathbf {w}_1^\star,\mathbf {\tilde w}_2,\mathbf {\tilde \Theta}) \leq R(\mathbf {w}_1^\star,\mathbf {w}_2^\star,\mathbf {\tilde \Theta})$, i.e., the objective value of problem \eqref{eq:SumRateMaxProb} increases after the first and second steps in each iteration of Algorithm~\ref{alg1}. Similarly, we can show that the objective value also increases after the third step in each iteration.}
Therefore, Algorithm~\ref{alg1} always converges. {Since problem \eqref{eq:SumRateMaxProb} is non-convex, Algorithm 1 cannot necessarily yield a global optimal solution. Nonetheless, simulation results in Section~\ref{sec:simu} show that it achieves excellent  performance under various scenarios.}

\end{appendices}

\end{document}